\documentclass[runningheads]{llncs}
\usepackage{cite}
\usepackage{subfig}

\usepackage{amsmath}
\usepackage{url}

\usepackage{enumitem}
\usepackage[font={small},labelfont=bf]{caption}

\usepackage{graphicx}
\usepackage{xcolor}
\usepackage{gensymb}

\newcommand{\wrapFigSpace}{\vspace{-5mm}}
\newcommand{\wrapTableSpace}{\vspace{-3mm}}
\usepackage[normalem]{ulem}

\begin{document}

%\title{Bare Demo of IEEEtran.cls\\ for IEEE Conferences}
\title{SRTGAN: Triplet Loss based Generative Adversarial Network for Real-World Super-Resolution}

% \author{\IEEEauthorblockN{Dhruv Patel\inst{\rm \thanks{* denotes equal contribution}*1} \and Abhinav Jain\inst{*1} \and Simran Bawkar\inst{1} \and Manav Khorasiya\inst{1} \and Kalpesh Prajapati\inst{1} \and Kishor Upla\inst{1} \and Kiran Raja\inst{2} \and Raghavendra Ramachandra\inst{2} \and Christoph Busch\inst{2}}}
\author{Dhruv Patel\inst{1}\thanks{denotes equal contribution} \and Abhinav Jain\inst{1}\textsuperscript{$\star$} \and Simran Bawkar\inst{1} \and Manav Khorasiya\inst{1} \and Kalpesh Prajapati\inst{1} \and Kishor Upla\inst{1} \and Kiran Raja\inst{2} \and Raghavendra Ramachandra\inst{2} \and Christoph Busch\inst{2}}
\authorrunning{D. Patel et al.}
\institute{
Sardar Vallabhbhai National Institute of Technology (SVNIT), Surat, India
\email{\{dhruv.r.patel14, abhinav98jain, sim017bawkar, manavkhorasiya, kalpesh.jp89, kishorupla\}@gmail.com}\\
\and
Norwegian University of Science and Technology (NTNU), Gjøvik, Norway.
\email{\{kiran.raja, raghavendra.ramachandra, christoph.busch\}@ntnu.no}}
% \thanks{$\star$ denotes equal contribution}


%Telephone: (800) 555--1212, Fax: (888) 555--1212}

\maketitle
\vspace{-0.5cm}
\begin{abstract}
Many applications such as forensics, surveillance, satellite imaging, medical imaging, etc., demand High-Resolution (HR) images. However, obtaining an HR image is not always possible due to the limitations of optical sensors and their costs. An alternative solution called Single Image Super-Resolution (SISR) is a software-driven approach that aims to take a Low-Resolution (LR) image and obtain the HR image. Most supervised SISR solutions use ground truth HR image as a target and do not include the information provided in the LR image, which could be valuable. In this work, we introduce Triplet Loss-based Generative Adversarial Network hereafter referred as \emph{SRTGAN} for Image Super-Resolution problem on real-world degradation. We introduce a new triplet-based adversarial loss function that exploits the information provided in the LR image by using it as a negative sample. Allowing the patch-based discriminator with access to both HR and LR images optimizes to better differentiate between HR and LR images; hence, improving the adversary. Further, we propose to fuse the adversarial loss, content loss, perceptual loss, and quality loss to obtain Super-Resolution (SR) image with high perceptual fidelity. We validate the superior performance of the proposed method over the other existing methods on the RealSR dataset in terms of quantitative and qualitative metrics.
%Obtaining a High Resolution(HR) image from a Low Resolution(LR) image has applications in surveillance, forensics, satellite imaging, medical imaging, etc. but obtaining an HR image is not always possible due to the limitations of optical sensors and their costs. Single Image Super-Resolution(SISR) aims to take an LR image and construct the HR image for the same. Most supervised SISR solutions make use of the HR image as a target and don't tend to include the information provided in the LR image which could be valuable. In this paper, we propose SRTGAN, a generative adversarial network (GAN) that uses triplet loss (T) to solve the problem of image super-resolution (SR). We introduce a new triplet-based adversarial loss function that tries to exploit the information provided in the LR image by using it as a negative sample. Allowing the patch-based discriminator access to both HR and LR images optimizes it to better differentiate between HR and LR images, hence improving the adversary. This adversarial loss along with other losses, namely content loss, perceptual loss, and quality loss pushes our solution to improve the perceptual fidelity of SR Images. We show on the realSR datasets competing results with the other state-of-the-art algorithms and an improved LPIPS score which reflects the human perceptual judgment. 
\end{abstract}

% \IEEEpeerreviewmaketitle

\section{Introduction}
% It is highly desirable to get High-Resolution (HR) image which contains more information with high precision in all application of computer vision. However, despite of large amount of work in super-resolution, it remains unsolved because of ill-posed nature which results in multiple solution from a single LR image. 

Single Image Super-Resolution (SISR) refers to reconstructing a High Resolution (HR) image from an input Low Resolution (LR) image. It has broad applications in various fields, including satellite imaging, medical imaging, forensics, security, robotics, where LR images are abundant. It is an inherently ill-posed problem since obtaining the SR image from an LR image might correspond to any patch of the ground truth HR image, which is intractable. The most employed solutions are the supervised super-resolution methods due to the availability of ground truth information and the development of many novel methods. % such as ESRGAN \cite{esrgan}.

\begin{figure*}[!t]
    \centering
    \includegraphics[width=0.45\linewidth]{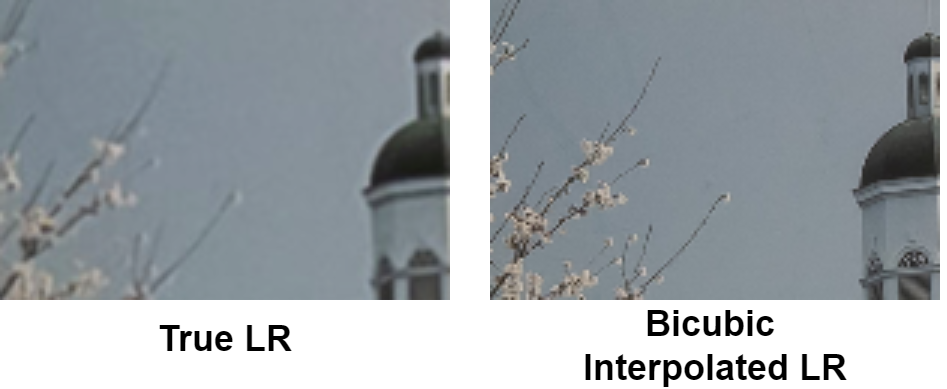}
  \hspace{0.3in}
    {\includegraphics[width=0.45\linewidth]{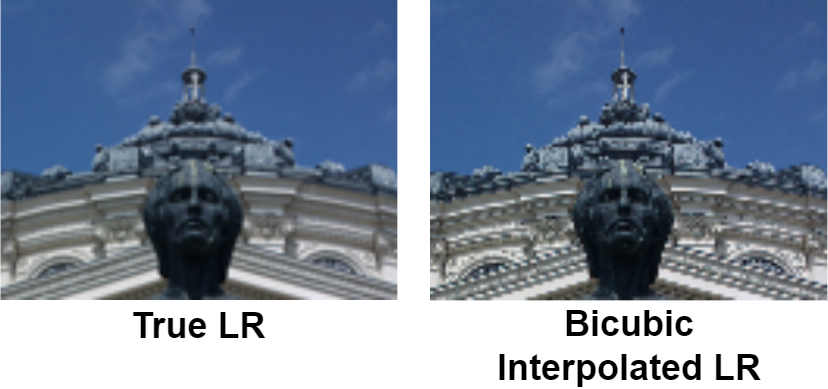}}\\
    \hspace{0.5in} RealSR dataset \cite{realsrdata} \hspace{1.0in} DIV2KRK Dataset \cite{KernelGAN}\\
        \caption{True LR and corresponding bicubic downsampled LR image from ground truth HR of the RealSR dataset \cite{realsrdata} and DIV2KRK dataset \cite{KernelGAN}}
        \wrapFigSpace
    \label{fig:qastat}
\end{figure*}

Reconstructing the HR image from LR input includes image deblurring, denoising, and super-resolution operations which makes the SISR a highly complex task. Due to recent technological advances, such as computational power and availability of data, there has been substantial development in various CNN architectures and loss functions to improve SISR methods \cite{SR-DRRN,SRCNN,srdensenet,FSRCNN,EDSR}. These models have been primarily tested on the synthetic datasets. Here, the LR images are downsampled from the ground truth HR images by using known degradation model such as bicubic downsampling. For instance, Fig.~\ref{fig:qastat} shows that the characteristics like blur and that details of true and bicubic downsampled LR images do not correspond exactly for both RealSR \cite{realsrdata} and DIV2KRK dataset \cite{KernelGAN}. Such differences can be attributed to underlying sensor noise and unknown real-world degradation. Hence, the models perform well on those synthetically degraded images, they generalize poorly on the real-world dataset \cite{blur_models}. Further, most of the works have shown that adding more CNN layers does increase the performance of the model by some extent. However, they are unable to capture the high-frequency information such as texture in the images as they rely on the pixel-wise losses and hence suffer from poor perceptual quality \cite{mathieu2016deep, vgg_loss, NIPS2016_371bce7d, bruna2016superresolution}. 

To address the issues mentioned above, the research community has also proposed using Generative Adversarial Networks (GANs) for SISR task. %These networks have been extensively used for image, audio, and video generation. 
%GAN based approaches are used to recover the final texture details at large upscaling factors. 
%The optimization-based super-resolution method is primarily driven by the choice of the objective function. However, they do not match the fidelity expectations in the high-resolution images  
The first GAN-based framework called SRGAN \cite{srgan}, introduced the concept of perceptual loss, calculated from high-level feature maps, and tried to solve the problem of poor perceptual fidelity as mentioned before. Subsequently, numerous GAN-based methods were introduced that have shown improvements in the super-resolution results \cite{esrgan,srgan,SRResCGAN}. 
%However, the state-of-the-art methods have shown performance improvements in either synthesized or real data. 
GANs are also used for generating perceptually better images \cite{esrgan,srgan,SRFeat}. Motivated by such works, we propose SR using Triplet loss-based GAN (SRTGAN) - a triplet loss-based patch GAN comprising a generator trained in a multi-loss setting with a patch-based discriminator. % The discriminator takes a combination of input LR, ground truth HR, and generated SR image as its input. The inherent formulation of the triplet loss implicitly forces the discriminator to penalize an LR foreground patch more than an LR background patch, which is missing when directly trained using vanilla GAN. In a vanilla GAN, we train the discriminator to classify an image as HR or LR, each patch of the image would be scored in quality, the caveat being, even in HR images, the background is blurred and could be considered lower quality which would cause spurious loss. We overcome this issue in triplet loss by introducing the LR and HR images as negative and positive samples, respectively, and the SR image as an anchor. 

Our proposed method - SRTGAN gains superior Peak Signal-to-Noise Ratio (PSNR) and competing Structural Similarity Index (SSIM) \cite{SSIM} values on the RealSR dataset (real-world degradation) \cite{realsrdata}, which still cannot be considered a valid metric as they fail to capture the perceptual features. Hence, we also evaluate our performance on the perceptual measure, i.e. Learned Perceptual Image Patch Similarity (LPIPS) \cite{LPIPS} score. Our SRTGAN outperforms the other state-of-the-art methods in the quantitative evaluation of LPIPS and visual performance on the RealSR dataset. It also provides superior LPIPS results on the DIV2KRK dataset \cite{KernelGAN} (synthetic degradation). All our experiments on both RealSR and DIV2KRK datasets are done for an upscaling factor of $\times4$. Even though DIV2KRK happens to be a synthetic dataset, it has a highly complex and unknown degradation model. Hence, our proposed method has been trained and validated on these datasets proving the generalizability on the real-world data. 

Our key contributions in this work can therefore be listed as:
\begin{itemize}[noitemsep, topsep=0pt]
    \item[$\bullet$] We propose a new triplet-based adversarial loss function that exploits the information provided in the LR image by using it as a negative sample as well as the HR image which is used as a positive sample.
    \item[$\bullet$] A patchGAN-based discriminator network is utilized that assists the defined triplet loss function to train the generator network.
    \item[$\bullet$] The proposed SR method is trained on a linear combination of losses, namely the content, multi-layer perceptual, triplet-based adversarial, and quality assessment. Such fusion of different loss functions leads to superior quantitative and subjective quality of SR results as illustrated in the results.
    \item[$\bullet$] Additionally, different experiments have been conducted in the ablation study to judge the potential of our proposed approach. The superiority of the proposed method over other novel SR works has been demonstrated from the undertaken quantitative and qualitative studies.
\end{itemize}
The structure of the paper is designed in the following manner. Section~\ref{sec:related} consists of the related work in the field. Section~\ref{sec:propose} includes the proposed framework, the network architecture, and loss formulation for training the Generator and Discriminator networks. The experimental validation is presented in Section~\ref{sec:exp}, followed by the limitation and conclusion in Section \ref{sec:limitations} and ~\ref{sec:conclusion} respectively.

\section{Related Works}
\label{sec:related}
\vspace{-1mm}
A Convolutional Neural Network (CNN) based SR approach (referred as SRCNN) was proposed by Dong et al. \cite{SRCNN}, where only three layers of convolution were used to correct finer details in an upsampled LR image. Similarly, FSRCNN \cite{FSRCNN} and VDSR \cite{jkim2016} were inspired by SRCNN with suitable modifications to further improve the performance. VDSR \cite{jkim2016} is the first model that uses a deep CNN and introduces the use of residual design that helps in the faster convergence with improvement in SR performance. Such residual connection also helps to avoid the vanishing gradient problem, which is the most common problem with deeper networks. Inspired by VDSR \cite{jkim2016}, several works \cite{RCAN1, EDSR, srgan, carnsr, usisrresnet} have been reported with the use of a residual connection to train deeper models. Apart from a residual network, an alternative approach using dense connections has been used to improve SR images in many recent networks \cite{srdensenet,rdn,dbpn}. The concept of attention was also used in several efforts \cite{RCAN1,pan} to focus on important features and allow sparse learning for the SR problem. Similarly, adversarial training \cite{gangoodfellow} has been shown to obtain better perceptual SR results. Ledig et al. introduced adversarial learning for super-resolution termed as SRGAN \cite{srgan}, which shows perceptual enhancement in the SR images even with low fidelity metrics such as PSNR and SSIM. Recent works such as SRFeat \cite{SRFeat} and ESRGAN \cite{esrgan}, which were inspired by SRGAN, have also reported improvements in the perceptual quality in obtaining SR images. A variant of GAN, TripletGAN \cite{tgan} demonstrated that a triplet loss setting will theoretically help the generator to converge to the given distribution. Inspired by TripletGAN, PGAN \cite{pgan} has been proposed, which uses triplet loss to super-resolve medical images in a multistage manner.\vspace{-4.5mm}

The limitation of the majority of the work mentioned above is the use of artificially degraded training data, such as bicubic downsampling. The CNNs typically fail to generalise well on the real-world data, because real-world degradation is considerably different than bicubic downsampling (see Fig.~\ref{fig:qastat}). The supervised approaches need real LR-HR pairs in order to generalise to real-world data, which is challenging. For recovering real-world HR images, Cai et al. \cite{realsrdata} introduced the RealSR dataset and a baseline network called Laplacian Pyramid-based Kernel Prediction Network (LP-KPN). Thereafter, several research works for SR have been conducted on the RealSR dataset, considering factors from real data into account. \cite{shiddnet,encoderntire2019,Feng_ntire2019,Gao_ntire2019,Du_ntire2019,Xu_ntire2019CVPR,Kwak_ntire2019}.\vspace{-4.5mm} 

Further, Cheng et al. suggested a residual network based on an encoder-decoder architecture for the real SR problem \cite{encoderntire2019}. A coarse-to-fine approach was used by them, where lost information was gradually recovered and the effects of noise were reduced. By adopting an autoencoder-based loss function, a fractal residual network was proposed by Kwak et al. \cite{Kwak_ntire2019} to super-resolve real-world LR images. At the outset of network architecture, an inverse pixel shuffle was also proposed by them to minimise the training parameters. Du et al. \cite{Du_ntire2019} suggested an Orientation-Aware Deep Neural Network(OA-DNN) for recovering of images with high fidelity. It is made up of many Orientation Attention Modules(OAMs) which are designed for extracting orientation-aware features in different directions. Additionally, Xu and Li have presented SCAN, a spatial colour attention-based network for real SR \cite{Xu_ntire2019CVPR}. Here, the attention module simultaneously exploits spectral and spatial dependencies present in colour images. In this direction, we provide a novel framework based on triplet loss in the manuscript inspired by \cite{tgan} to enhance the perceptual quality of SR images on the realSR dataset. \looseness=-1\vspace{-3.35mm}

Although there have been previous attempts to incorporate the triplet loss optimization for super-resolution such as PGAN \cite{pgan}, which progressively super-resolve the images in a multistage manner, it has to be noted that they are specifically targeted to medical images, and in addition, the LR images used are obtained through a known degradation (such as bicubic sampling) and blurring (Gaussian filtering). Thus, it fails to address real-world degradation. Using the triplet loss, the proposed patch-based discriminator can better distinguish between generated and high-resolution images, thereby improving the perceptual fidelity. To the best of our knowledge, the utilization of triplet loss to the real-world SISR problem has not been explored before. We, therefore, propose the new approach as explained in the upcoming section.

\section{Proposed Method}
\label{sec:propose}
\begin{figure}[h!]
    \centering
    \includegraphics[height=0.55\linewidth,width=0.80\linewidth]{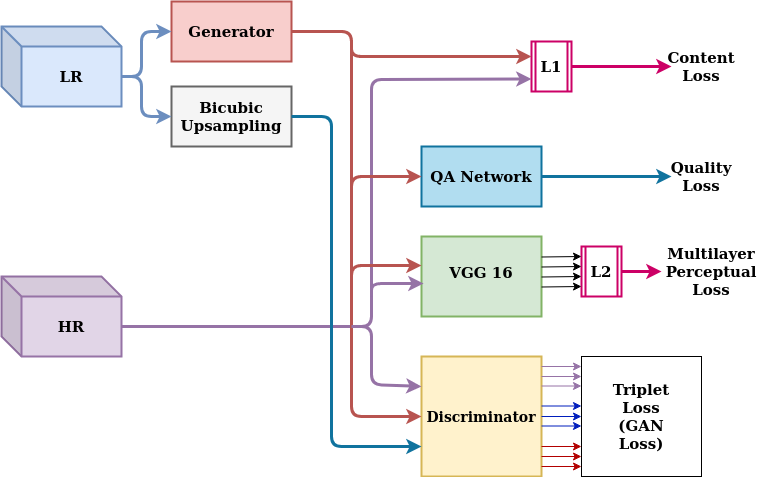}
    \caption{The training framework of our proposed method - SRTGAN.}
    \label{fig:network}
    \wrapFigSpace
\end{figure}

\vspace{-0.5mm}
Fig. \ref{fig:network} shows the detailed training framework of our proposed method. The proposed supervised SR method expects the LR and its corresponding ground truth HR image as the input. It performs super-resolution on the LR image using the generator network, which is trained in a multi-loss setting using a fusion of losses namely content, perceptual, adversarial, and quality assessment. As depicted in Fig.~\ref{fig:network}, the content Loss is calculated as $L_1$ loss (pixel-based difference) between the generated(SR) and ground truth(HR) images. It assists the generator in preserving the content of ground truth HR. As the generator network is trained in an adversarial setting with the discriminator, we use a triplet-based GAN loss, which also boosts the stability of the learning. Apart from the GAN loss, we incorporate multi-layer perceptual loss, which is calculated as $L_2$ loss between the features of HR and SR, obtained from a pre-trained VGG network as suggested in SRGAN \cite{srgan}. Moreover, we also use a quality assessment loss based on Mean Opinion Score (MOS) for improving the perceptual quality of generated images \cite{usisrresnet}. The validation of each setting in the framework is demonstrated in the ablation section later. \looseness=-1

% 1.) Generator  Network

% 2.) QA Network

% 3.) Discriminator Network

%  The overall proposed architecture can be seen in Fig. \ref{fig:network}. Generator produces an SR image by taking LR image as input. Content loss(L1 based) is calculated between generated SR image and corresponding HR output. This content loss helps the Generator in preserving the structure of LR input in the generated SR image. In addition, we introduce a Quality Assessment (QA) network to function on the generated SR image and produce a quality score for that SR image. This quality score is based on human perception and is used as a loss function for the Generator, which helps it to improve the quality of SR image.

\begin{figure}[t!]
    \centering
    \includegraphics[height=0.55\linewidth,width=0.80\linewidth]{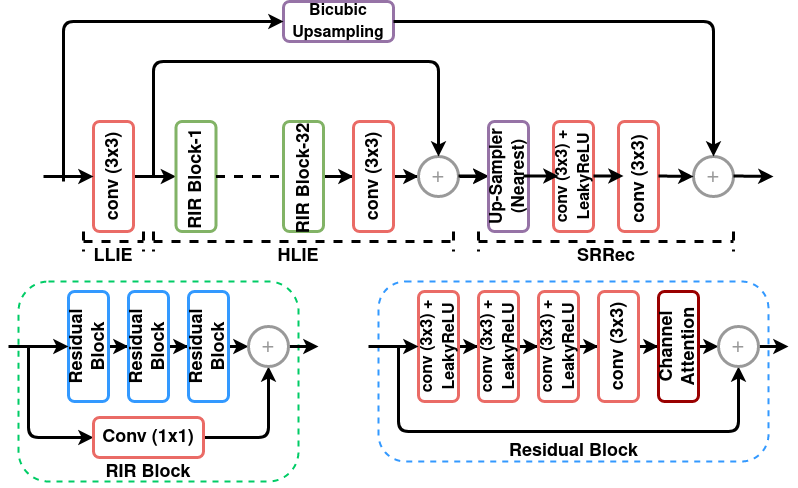}
    \caption{Generator network \cite{icpr_paper}.}
    \label{fig:gen}
    \wrapFigSpace
\end{figure}

\textbf{Generator Network (G):} The design of generator network is shown in Fig. \ref{fig:gen}, which was published in \cite{icpr_paper}. The architecture can be divided into Feature Extraction (Low-level Information Extraction (LLIE), High-level Information Extraction (HLIE)) and Reconstruction (SR reconstruction (SRRec)) modules based on their functionality.
% The architecture can be divided into Low-Level Information Extraction (LLIE), High-Level Information Extraction (HLIE), and SR reconstruction (SRRec) modules based on its functionality. 
The LLIE module is initially fed with LR input ($I_{LR}$) for extracting the low-level details $(i.e., I_{l})$. It consists of a convolutional layer with kernel size $3$ and $32$ channels.
% Larger kernel is used, which leads to larger reception area for predicting the accurate low-level information.
This can be expressed mathematically as,
\begin{equation}
    I_{l} = f_{LLIE}(I_{LR}).
\end{equation}

The edges and fine structural details present in the LR image are extracted by the HLIE module using the low-level information $I_{l}$. HLIE module comprises of 32 Residual-In-Residual (RIR) blocks, one $3 \times 3$ convolutional layer, and have one long skip connection. The long skip connection here stabilizes the network training \cite{srgan, usisrresnet, esrgan, icpr_paper}. Each RIR block is created using three residual blocks and a skip connection with a $1 \times 1$ convolutional layer. The Residual Block comprises of four $3 \times 3 $ convolutional layers with a serially attached Channel Attention (CA) module. Using the statistical average of each channel, each channel is independently re-scaled via the CA module \cite{RCAN1}. As depicted in Fig.~\ref{fig:gen}, skip connections are also used in residual blocks, which aids in stabilizing the training of deeper networks and resolving the vanishing gradient problem. The output from HLIE module can be expressed as,
\begin{equation}
    I_{h} = f_{HLIE}(I_l).
\end{equation}
 Now, feature maps with high-level information (i.e. $I_{h}$) are passed to the SR Reconstruction (SRRec) module, which comprises of 1 up-sampling block and 2 convolutional layers. This helps in mapping $I_{h}$ to the required number of channels needed for output image $(I_{SR})$. This can be stated as follows:
\begin{equation}
    I_{SR} = f_{REC}(I_{h}),
\end{equation}
where the reconstruction function of the SRRec module is $f_{REC}$. The nearest neighbour is used to perform a $2\times$ upsampling with a $3 \times 3$ convolutional layer and 32 feature maps in each up-sampling block. Finally, a convolutional layer is used to map 32 channels into 3 channels of SR image in the generator network.

\begin{figure}[!h]
    \centering
    \includegraphics[height=0.25\linewidth,width=0.60\linewidth]{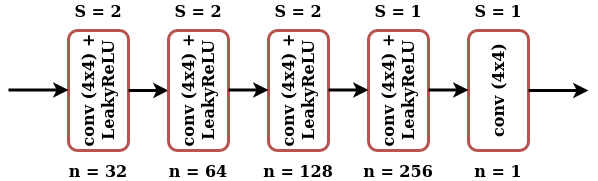}
    \caption{Discriminator Network. Here, $n$ stands for the number of channels, while $S$ represents stride.}
    \label{fig:disc}
    \wrapFigSpace
    \vspace{-1mm}
\end{figure}

\textbf{Discriminator (D) Network:}
%\vspace{-2mm}
We further use a PatchGAN \cite{patchgan} based discriminator network to distinguish foreground and background on a patch with scale of $70 \times 70$ pixels. The proposed architecture is shown in Fig.~\ref{fig:disc}. It is designed by adhering to the recommendations made in the work of PatchGAN \cite{patchgan}. It consists of five convolutional layers with strided convolutions. After each convolution, the number of channels doubles, excluding the last output layer which has a single channel. The network uses a fixed stride of two except for the second last and last layer where the stride is set to 1. It is noted that a fixed kernel size of $4$ is used for all layers throughout the discriminator network. 
Further, each convolutional layer except the output layer uses leaky ReLU activation and padding of size one. All intermediate convolutional layers except the first and last layer use Batch Normalisation.

\begin{figure}[!h]
    \centering
    \includegraphics[height=0.35\linewidth,width=1.0\linewidth]{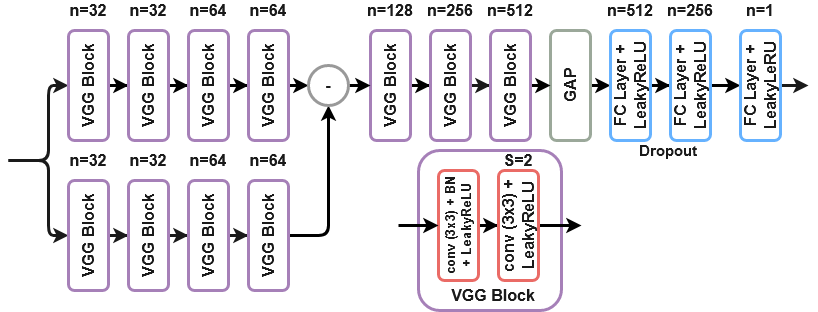}
    \caption{The architecture of QA network \cite{icpr_paper}.}
    \label{fig:qa}
    \wrapFigSpace
    \wrapTableSpace
\end{figure}

\textbf{Quality Assessment (QA) Network:} 
Inspired by \cite{icpr_paper}, a novel quality-based score obtained from QA Network is employed which serves as a loss function in training. The design of QA network is shown in Fig. \ref{fig:qa}, which is inspired by the VGG. The addition of the QA loss in the overall optimization enhances the image quality based on human perception as the QA network is trained to mimic how humans rank images based on their quality. Instead of using a single path to feed input to the network, two paths have been employed in this case. To proceed forward, both of these features are subtracted. Each VGG block has two convolutional layers, the second of which uses a stride of $2$ to reduce the spatial dimensions. The network uses Global Average Pooling (GAP) layer instead of flattening layer to minimize the trainable parameters. At fully connected layers, a drop-out technique is used to overcome the issue of over-fitting. The KADID-10K \cite{kadid} dataset, consisting of $10,050$ images, was used to train the QA network. The dataset has been divided in 70\%-10\%-20\% ratio for train-validate-test purposes respectively during the training process. 
\subsection{Loss Functions}
As depicted in Fig.~\ref{fig:network}, the generator is trained using a fusion of content loss (pixel-wise $L_1$ loss), GAN loss (triplet-based), QA loss, and perceptual loss. Mathematically, we can describe the loss of generator by the following formula:
\begin{equation}
    \label{eqn:gen}
     L^{gen} = \lambda_1L_{content} + \lambda_2L_{QA} + \lambda_3L_{GAN}^G + \lambda_4L_{perceptual}.
\end{equation}
The values of $\lambda_1$, $\lambda_2$, $\lambda_3$ and $\lambda_4$ are set empirically to $5$, $2\times 10^{-7}$, $1\times10^{-1}$ and $5\times10^{-1}$, respectively. 

The Discriminator network is trained using triplet-based GAN loss. This can be expressed as,
\begin{equation}
     L^{disc} = \lambda_3L_{GAN}^D 
 \end{equation}
where $\lambda_3$ is emperically set to $1\times10^{-1}$.

Both  $L_{GAN}^G$ and $ L_{GAN}^D$ are defined in Eqn.~\ref{eqn:gen_gan} and \ref{eqn:dis_gan} respectively.

The content loss in Eqn. \ref{eqn:gen} has been used to preserve the content of the ground truth, which is an $L_1$ loss between ground truth HR (i.e., $I_{HR}$) and generated image SR (i.e., $I_{SR}$), and same can be expressed as,
% \begin{equation}
%     L_{content} = \sum^{N}\| I^{SR} - I^{up}\|_1
% \end{equation}
\begin{equation}
    L_{content} = \sum^{N}\| G(I_{LR}) - I_{HR})\|_1, 
\end{equation}
where $N$ denotes the batch size in training, and $G$ represents the function of generator. $\sum^N[\cdot]$ denotes an average operation across all images in the mini-batch. The perceptual loss $L_{perceptual}$ is used here for improving the perceptual similarity of the generated image with respect to its ground truth, which can be expressed as,
\begin{equation}
    L_{perceptual} = \sum^N\big[\sum^4_{i=1}MSE(F_{HR}^i,F_{SR}^i)\big].
\end{equation}
Here, $MSE(a,b)$ represents Mean Square Error (MSE) between $a$ and $b$, $F^i$: Normalised features taken from $layers[i]$ and $layers$ = [$relu_{12}$, $relu_{22}$, $relu_{33}$, $relu_{43}$]. Here, $layers$ is the list of four layers of VGG-16 used for the calculation of perceptual loss \cite{vgg_loss}. Such loss is calculated as the MSE between the normalized feature representations of generated image ($F_{SR}$) and ground truth HR ($F_{HR}$) obtained from a pre-trained VGG-16 network. It is not dependent on low-level per-pixel information that leads to blurry results. Instead, it depends on the difference in high-level feature representations which helps to generate images of high perceptual quality. In addition, the idea of using multi-layer feature representations adds to its robustness.
To further improve the quality of SR images based on human perception, a Quality Assessment (QA) loss is also introduced. It rates the SR image on a scale of 1-5, with a higher value indicating better quality. This predicted value is used to calculate the QA loss i.e., $L_{QA}$, which is expressed as \cite{icpr_paper},
\begin{equation}
    L_{QA} = \sum^N\big(5 - Q(I_{SR})\big),
\end{equation}
where $Q(I_{SR})$ represents the quality score of SR image from the QA network.
% \begin{figure*}[t!]
%     \centering
%     \includegraphics[width=0.95\textwidth]{Qualitative Analysis of Proposed method on RealSR modified.png}
%     \caption{The comparison of the Super-Resolution results obtained using the Proposed Architecture and Without Quality Loss method and Without Triplet Loss (Vanilla GAN Loss) method on RealSR validation dataset \cite{realsrdata}.}
%     \label{fig:val1}
% \end{figure*}

% \begin{figure*}[t!]
%     \centering
%     \includegraphics[width=0.95\textwidth]{Qualitative Analysis of Proposed method on DIV2KRK_9 modified.png}
%     \caption{The comparison of the Super-Resolution results obtained using the Proposed Architecture and Without Quality Loss method and Without Triplet Loss (Vanilla GAN Loss) method on DIV2KRK dataset \cite{div2kcite}}
%     \label{fig:val2}   
% \end{figure*}

% \begin{center}
%     \centering
%     \begin{tabular}{|c|c|c|c|}
%         \hline
%         $relu_1_2$ & $relu_2_2$ & $relu_3_3$ & $relu_4_3$ \\
%         \hline
%     \end{tabular}\\
% \end{center}

The GAN loss used here is a triplet-based loss function to a patch-based discriminator. 
% /* random stuff will need to find citations
% Why patchgan - coz patches rock - cite paper */
\begin{figure*}[t!]
    \centering
    \includegraphics[width=0.95\textwidth]{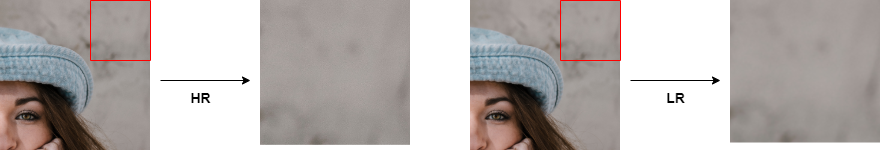}
    \caption{Comparison of background patch in LR and HR images.}
    \label{fig:bghrlrcomp}
    \wrapFigSpace
\end{figure*}
An image can be simplified consisting of 2 parts, Background and Foreground; according to human perceptions, we rate images to be higher quality based on the foreground, which is the focus of the image. On the other hand, the background between LR and HR images is hard to differentiate as shown in Fig \ref{fig:bghrlrcomp}. 
% \dhruv{When using a Vanilla Patch GAN with a mean square loss on a background patch, the output of the discriminator will be 0.5 in an ideal case as the patch would have no information to be rated on its quality. The losses for both the discriminator and generator for this case will be as follows. G is the Generator function, D is the discriminator function
% \begin{eqnarray}
%       L_D = D(I_{HR})^2 + (1-D(G(I_{LR}))^2 \nonumber \\
%       L_G = (1-D(G(I_{LR}))^2 \nonumber \\
%       D(I_{HR}) = D(G(I_{LR})) = 0.5   \nonumber \\
%       \therefore L_D = 0.5, L_G = 0.25\nonumber 
% \end{eqnarray}
% }  
A background patch with a vanilla GAN would be similar to the discriminator perceptually, hence forcing the output for the same to be real/fake could lead to a high erroneous loss and cause instability and noise in training. However, in the case of foreground patches, the idea of vanilla GAN will work well. To solve this problem, we introduce the use of triplet loss: instead of forcing the discriminator output for HR and SR to be opposite labels, we calculate the loss using the relative output produced by the discriminator for HR, LR, and SR images. We formulate this as triplet loss optimization comprising of 3 variables - positive, negative and anchor. The distance between the anchor and the positive is minimised by the cost function, while the distance between the anchor and the negative is maximised. For the generator, the anchor is defined as the generated SR image ($I_{SR}$), the positive as the ground-truth HR image ($I_{HR}$), and the negative as the up-sampled LR input ($n$($I_{LR}$)), where $n$ is the bicubic upsampling factor. The positive and negative are interchanged for training the discriminator.
 Thus, the triplet-based GAN losses for generator and discriminator can be defined as,
{\footnotesize
\begin{eqnarray}
    L_{GAN}^G =  \sum^N\big[MSE(D(I_{SR}), D(I_{HR})) - MSE(D(I_{SR}), D(n(I_{LR}))) + 1\big] \label{eqn:gen_gan}\\
    L_{GAN}^D =  \sum^N\big[MSE(D(I_{SR}), D(n(I_{LR}))) - MSE(D(I_{SR}), D(I_{HR})) + 1\big] \label{eqn:dis_gan}
\end{eqnarray}}
Here, $MSE(a,b)$ represents mean square error between $a$ and $b$; $n$ denotes upsampling factor. 
% \dhruv{To compare to a vanilla GAN we again calculate generator and discriminator losses for a background patch
% \begin{eqnarray}
%       D(I_{HR}) = D(I_{SR}) = D(n(I_{LR})) = 0.5   \nonumber \\
%       \therefore L_{D_{triplet}} = (0.5-0.5)^2 - (0.5-0.5)^2= 0\nonumber \\
%                 L_{G_{triplet}} = (0.5-0.5)^2 - (0.5-0.5)^2= 0\nonumber 
% \end{eqnarray}
% As we see for a background patch, no error is propagated when using a triplet loss.} 
This triplet based GAN loss teaches the Generator to generate sharp and high-resolution images by trying to converge SR embeddings $D(I_{SR})$ and HR embeddings $D(I_{HR})$ and diverge SR embeddings with LR embeddings $D(n(I_{LR}))$, which are obtained from the Discriminator. Simultaneously, it also trains the patch-based Discriminator to distinguish the generated SR image from the ground-truth HR. The background patch as discussed before is similar for LR and HR images. Applying this triplet-based GAN loss patch-wise, improves the adversary as it allows the discriminator to better distinguish the main subject(foreground) of SR and HR images, which helps in generating images with better perceptual fidelity. 
%\ku{In the figure of proposed framework, we are saying L1 distance for domain loss and here we are saying absolute distance, which is correct? Sir L1 norm is same as absolute summation as I am thinking. But to match notation with eqn 7, I change it in L1 norm representation.}
\vspace{-2mm}
\section{Experimental Results}
\label{sec:exp}
\vspace{-2mm}
%\kr{Please check this for consistency on datasets, I have made some changes.}
%All experiments have been conducted on a computer with Intel Xeon(R) CPU with 128GB RAM and NVIDIA Quadro P5000 GPU with 16GB memory. Hyper-parameter tuning, visual and quantitative evaluations of the proposed approach with other state-of-the-art methods have been elaborated in the following subsections.
\subsection{Training Details}
\vspace{-0.5mm}
Using our proposed framework, we conduct supervised training on the RealSR dataset \cite{realsrdata}. In this dataset, the focal length of a digital camera has been adjusted to collect LR-HR pairs of the same scene. To incrementally align the image pairs at various resolutions, an image registration method is developed. Our proposed network has been trained on 400 such images from the RealSR dataset and additionally it has been validated on 100 LR-HR image pairs provided in the same dataset. Finally, DIV2KRK \cite{KernelGAN} and test set of RealSR dataset \cite{realsrdata} are employed for testing purposes.
%100 number of LR-HR image pairs from DIV2K dataset \cite{div2krkcite} had been employed as provided by organizers of NTIRE-2020 Real-world SR Challenge \cite{NTIRE2020RWSRchallenge}.
% The artificial degradation has been carried out, enabling us to measure the performance in terms of PSNR or SSIM values \textcolor{red}{FOR WHICH DATASET.. WRITE HERE}. 
The LR images are subjected to several augmentations during the training phase, including horizontal flipping, rotation of 0${\degree}$ or 90${\degree}$, and cropping operations.% We train our model using Adam optimizer up to $1,500$ iterations with a batch size of $32$. We keep $\beta_1$ value as 0.9, and set learning rate at $1\times10^{-5}$. We decrease this learning rate by half after every $500$ iterations. 
 The total trainable parameters of generator and discriminator networks are $3.7M$ and $2.7M$, respectively.  

Additionally, we also employ QA network-based loss to enhance the quality of generated images. This method has been referenced from the work of \cite{icpr_paper}. Our proposed triplet loss optimization improves the visual appearance of the SR images to make them more realistic. 
% \begin{figure}[!t]
%     \centering
%     \includegraphics[height=0.6\linewidth,width=0.95\linewidth]{KADID_Result_VGGGAP.png}
%     \caption{Result of QA network \cite{icpr_paper}: Actual MOS vs predicted MOS on KADID-10K \cite{kadid} testing dataset}
%     \label{fig:qa_result}
% \end{figure}
\vspace{-1mm}
\subsection{Ablation Study}

\begin{figure*}[t!]
    \centering
    \includegraphics[width=0.95\textwidth]{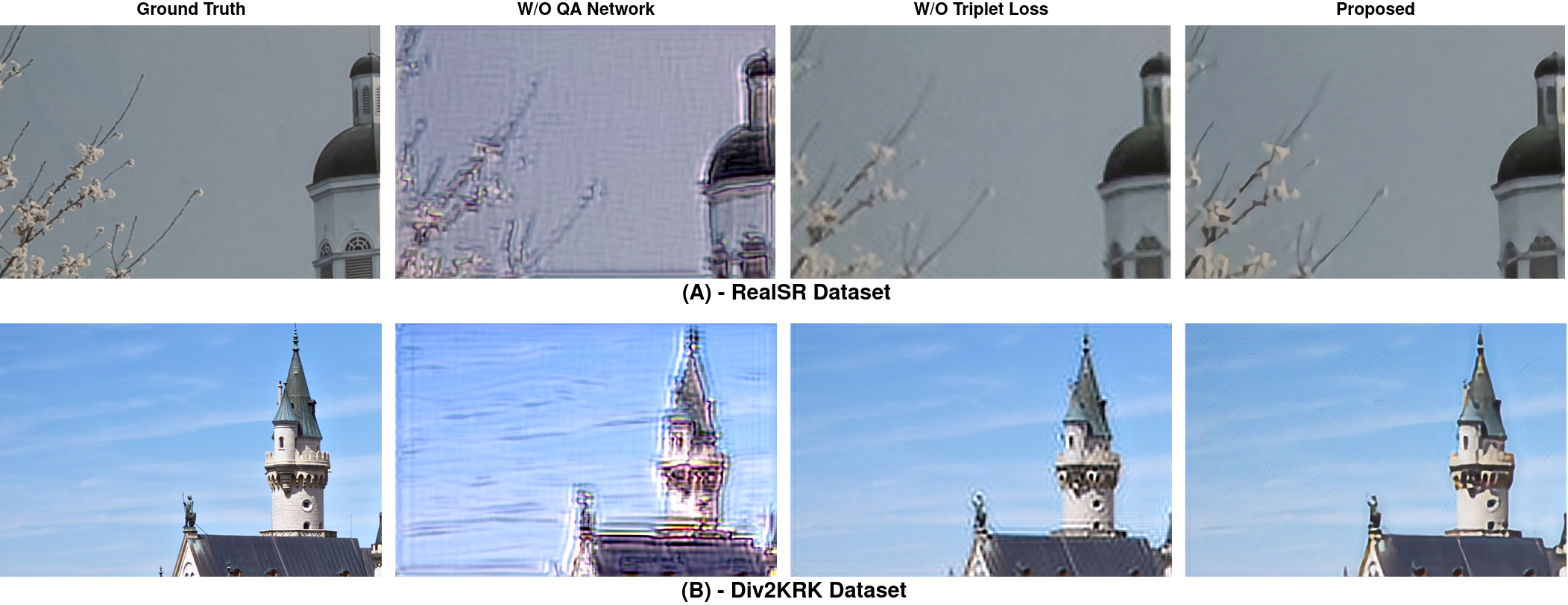}
    \caption{Comparison of the results obtained through our proposed method-\emph{SRTGAN} (with QA network and Triplet loss) Vs without incorporating QA Network or Triplet Loss on (A)-RealSR dataset \cite{realsrdata} and (B)-DIV2KRK dataset \cite{KernelGAN}}
    \label{fig:val1}
    \wrapTableSpace
\end{figure*}

\begin{table}[t!]
\centering
\caption{Quantitative evaluation of SRTGAN (with QA Network and Triplet Loss) Vs without incorporating these modules on the RealSR dataset \cite{realsrdata}.}
\resizebox{\linewidth}{!}{
%\resizebox{0.5\textwidth}{!}{\input{tables/table1}}
%\resizebox{0.80\textwidth}{!}{
\begin{tabular}{lccc} 
\hline
\textbf{Method} & \textbf{PSNR} \textuparrow      & \textbf{SSIM} \cite{SSIM}\textuparrow           & \textbf{LPIPS} \cite{LPIPS}\textdownarrow          \\ 
\hline
w/o Triplet Loss (Vanilla GAN Loss) & 25.879 & 0.72199          & 0.37095     \\   %We are yet to discuss values with sir 

w/o QA Network    &   16.126 &  0.39542  &  0.51217 \\
\textbf{Proposed} & \textbf{26.47283} & \textbf{0.754585} & \textbf{0.283878}\\
\hline
\end{tabular}}
\label{tab:abl1}
\wrapTableSpace
\end{table}

% \begin{figure}[!t]
%     \centering
%     \subfloat[w/o Disc.]{
%     \includegraphics[width=0.11\textwidth]{a_wo_disc.png}}
%     \hspace{0.2pt}
%     \subfloat[w/o QA loss]{\includegraphics[width=0.11\textwidth]{a_wo_qa.png}}
%     \hspace{0.2pt}
%     \subfloat[proposed]{\includegraphics[width=0.11\textwidth]{a_proposed.png}}
%     \hspace{0.2pt}
%     \subfloat[Ground-truth]{\includegraphics[width=0.11\textwidth]{a_hr.png}}
%     \caption{The SR results obtained using the proposed method (a) without discriminator network and (b) without QA network. (c) Proposed method and (d) ground-truth HR image. (better visualization in zoomed images)}
%     \label{fig:ablation}
%     %\vspace{-5pt}
% \end{figure}
%In the proposed method, we employ discriminator network in addition to VAE and QA networks to improve the quality of SR images.
\vspace{-0.5mm}
We demonstrate the experimental support for incorporating the triplet loss and QA network in this section. Quantitative and Qualitative assessment conducted on the RealSR dataset \cite{realsrdata}, are shown in Table~\ref{tab:abl1} and Fig.~\ref{fig:val1}, respectively. Our method yields superior SR outcomes on both synthetic and real-world data (RealSR dataset). The proposed method with QA network and Triplet Loss performs better (see Table~\ref{tab:abl1}) when compared to the performance obtained using the framework without those modules. This is quantitatively evaluated on various distortion metrics like PSNR and SSIM and perceptual measures, such as LPIPS. %shows the SR results obtained using such ablation study. In
The SR images produced using our proposed approach with QA network and Triplet Loss are also perceptually better when compared to without adding these modules, which is shown in Fig.~\ref{fig:val1}. It has been observed that our method without QA Network generates blurry output and variation in the natural color of the image. Our framework when optimized using vanilla GAN loss(instead of triplet loss), closely resembles the colour as anticipated in the real world, but fails to sharpen the edges, causing blurring. The proposed method's advantage may be observed in its ability to produce SR images with an adequate level of sharpening around the edges and preserving the color-coding of the original image. Here, by observing Fig.~\ref{fig:val1}, one may quickly determine the perceptual improvement from our proposed strategy.

\wrapTableSpace
\subsection{Quantitative Analysis}
\label{subsec:quantitative}
\vspace{-1mm}
The PSNR and SSIM values, which are the accepted measurements for the SR problem, are often estimated for comparison of the results between different approaches. These metrics, however, do not entirely justify the quality based on human perception. Therefore, we also estimate a full-reference perceptual quality assessment score known as LPIPS \cite{LPIPS}. A low LPIPS score indicates a better visual quality.

\begin{table}[!t]
    \centering
        \caption{Quantitative evaluation of SRTGAN with other state-of-the-art SR methods on RealSR and DIV2KRK dataset} \label{tab:Compare}
        \resizebox{\linewidth}{!}
        {%
        \setlength\tabcolsep{0.5pt} %to shrink left and right padding in cell(default value is 6pt)
    \begin{tabular}{|l|c|c|c|c|c|c|}
        \hline
        \textbf{Method} & \textbf{PSNR} \textuparrow & \textbf{SSIM} \cite{SSIM} \textuparrow & \textbf{LPIPS} \cite{LPIPS} \textdownarrow & \textbf{PSNR} \textuparrow & \textbf{SSIM} \cite{SSIM} \textuparrow & \textbf{LPIPS} \cite{LPIPS} \textdownarrow   \\
        \hline 
         & \multicolumn{3}{|c|}{\textit{DIV2KRK \cite{KernelGAN} Dataset}} & \multicolumn{3}{|c|}{\textit{ RealSR \cite{realsrdata} Dataset}} \\ 
        \hline 
        Bicubic & 23.89 & 0.6478 & 0.5645 & 25.74 & 0.7413 & 0.4666 \\
        \hline
        ZSSR \cite{zeroshotsr} & 24.05 & 0.6550 & 0.5257 & 25.83 & 0.7434 & 0.3503 \\
        \hline
        KernelGAN \cite{KernelGAN} & 24.76 & 0.6799 & 0.4980 & 24.09 & 0.7243 & 0.2981 \\
        \hline
        DBPI \cite{DBPI} & 24.92 & 0.7035 & 0.4039 & 22.36 & 0.6562 & 0.3106 \\
        \hline
        DAN \cite{DAN} & \textbf{26.07} & \textbf{0.7305} & 0.4045 & 26.20 & \textbf{0.7598} & 0.4095 \\
        \hline
        IKC \cite{IKC} & 25.41 & 0.7255 & 0.3977 & 25.60 & 0.7488 & 0.3188 \\
        \hline
        SRResCGAN \cite{SRResCGAN} & 24.00 & 0.6497 & 0.5054 & 25.84 & 0.7459 & 0.3746 \\
        \hline
        \textbf{Proposed} & 24.17 & 0.6956 & \textbf{0.3341} & \textbf{26.47} &  0.7546 & \textbf{0.2838} \\
        \hline
    \end{tabular}}
\wrapTableSpace
\end{table}

The comparison of all three metrics on the DIV2KRK \cite{KernelGAN} and RealSR datasets \cite{realsrdata} is presented in Table \ref{tab:Compare}. On both datasets, SRTGAN outperforms other novel approaches on LPIPS metric, demonstrating the proposed method's superiority in terms of perceptual quality. Our proposed approach also performs superior to other methods on PSNR metric, whereas performs competitively in terms of SSIM, on the RealSR dataset \cite{realsrdata}. SRTGAN also performs quite competitively in terms of PSNR and SSIM on the synthetic dataset - DIV2KRK \cite{KernelGAN}. The perceptual metric, LPIPS obtained using our proposed approach is significantly better for both datasets (see Table~\ref{tab:Compare}). 
\vspace{-1mm}
\begin{figure*}[t!]
    \centering
    
    \subfloat[Results on RealSR dataset \cite{realsrdata}.\label{fig:val3}]{\includegraphics[width=0.95\textwidth]{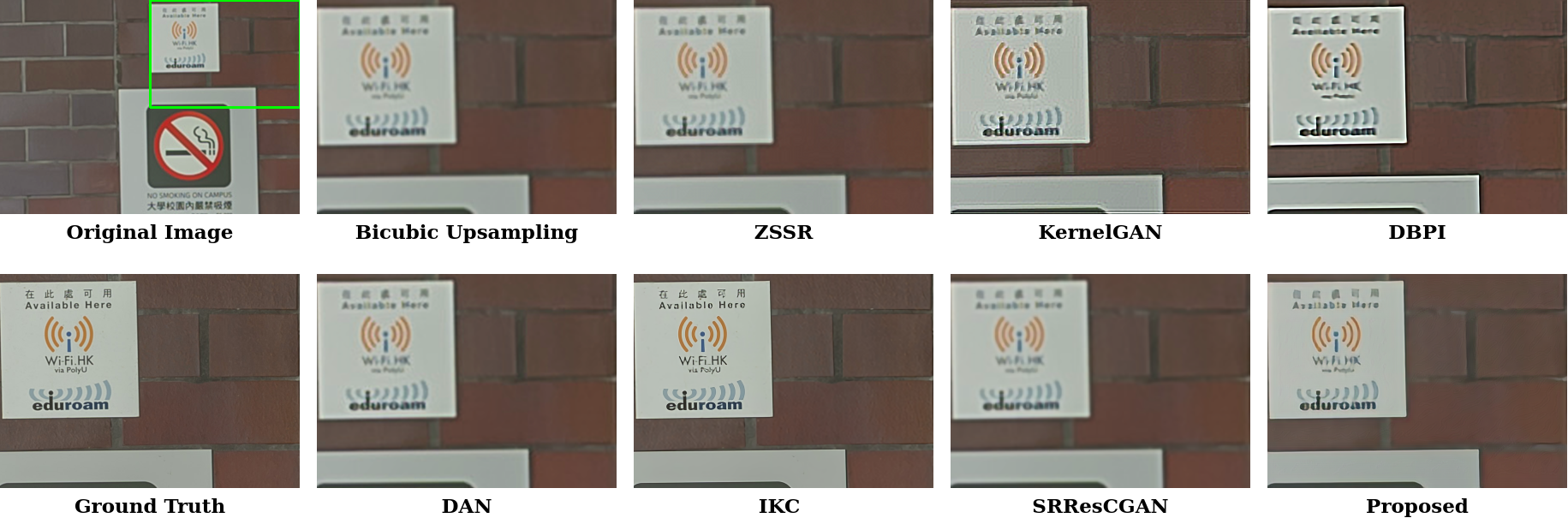}}
    % \hfil
    \vspace{-0.5mm}
    \subfloat[Results on DIV2KRK dataset \cite{KernelGAN}.\label{fig:val4}]{\includegraphics[width=0.95\textwidth]{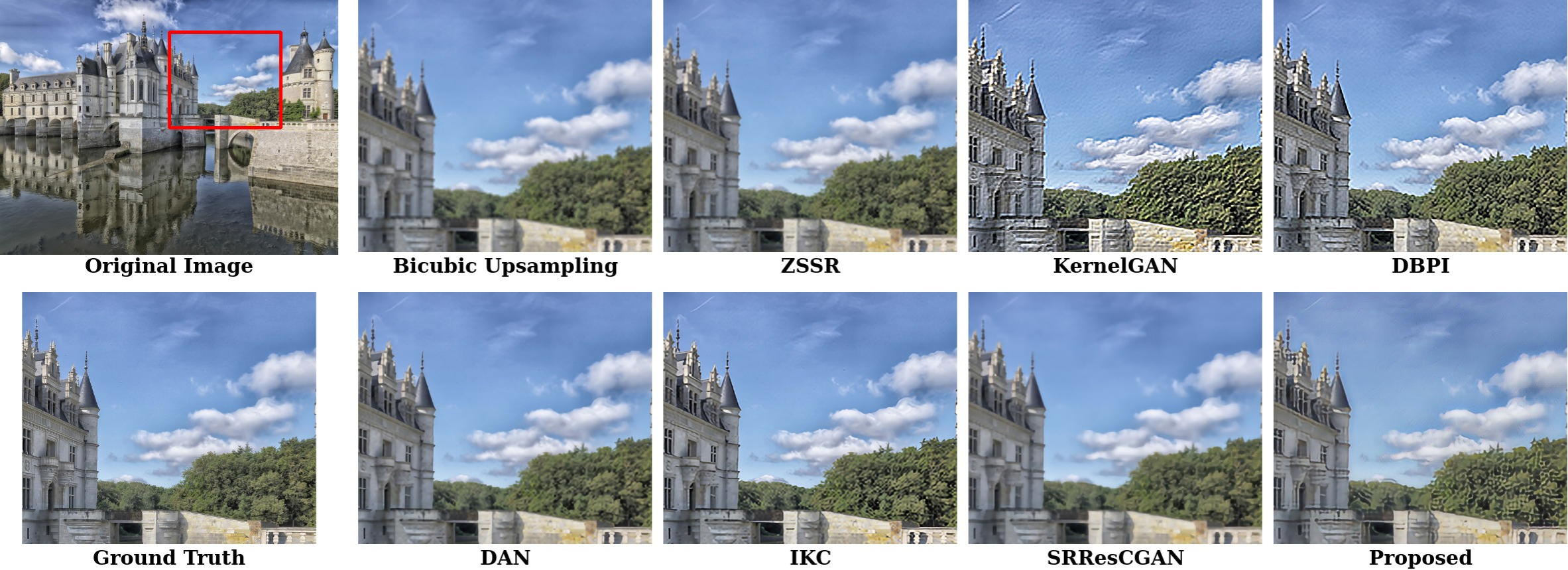}}
    % \hfil
    \vspace{-0.5mm}
    \subfloat[Results on DIV2KRK dataset \cite{KernelGAN}.\label{fig:val5}]{\includegraphics[width=0.95\textwidth]{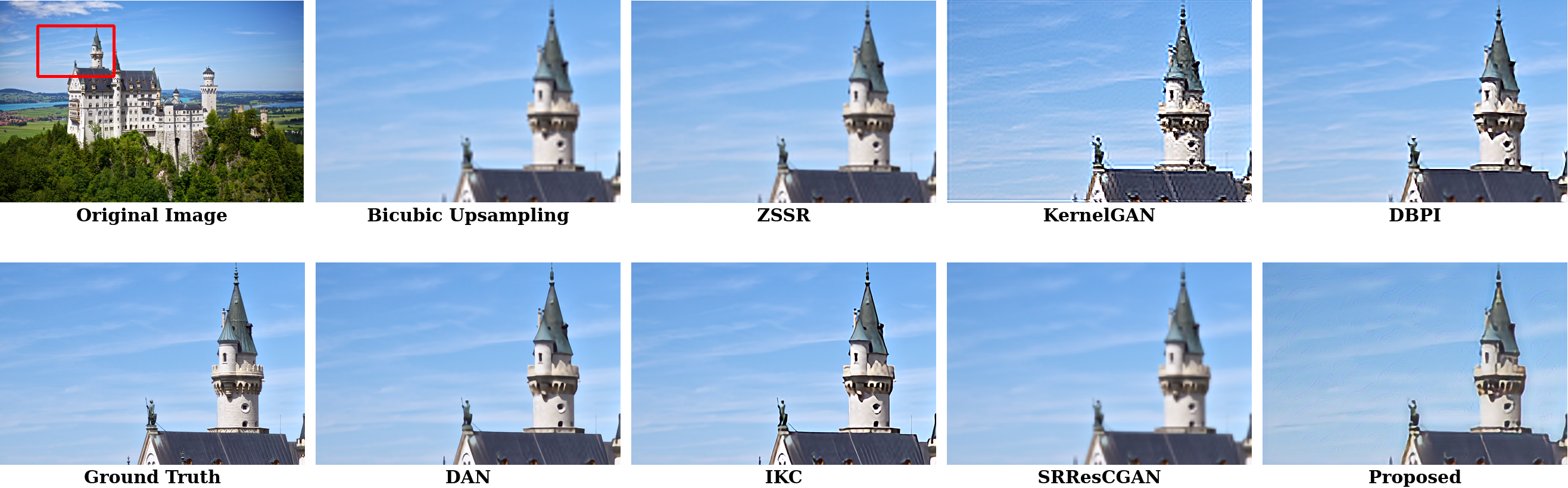}}
    \vspace{-3mm}
    \caption{Qualitative evaluation of SRTGAN with other state-of-the-art methods on RealSR and DIV2KRK dataset}
    \label{fig:val}
    \wrapTableSpace
\end{figure*}

\subsection{Qualitative Analysis}
\label{subsec:qual}
\vspace{-1mm}
In this section, we show the efficacy of SRTGAN through visual inspection. We qualitatively evaluate the SR performance on one image of RealSR dataset (Fig.~\ref{fig:val3}) \cite{realsrdata} and two sample images of DIV2KRK dataset (Fig.~\ref{fig:val4} and ~\ref{fig:val5}) \cite{KernelGAN}. %We compare the SR results on an image of RealSR validation dataset \cite{realsrdata} in which original HR image (Ground Truth) is available as shown in Fig. \ref{fig:val3}. Additionally, we have also performed a comparison on DIV2KRK dataset \cite{div2krkcite} in Fig.~\ref{fig:val4} and Fig.~\ref{fig:val5}.
In addition, we also make comparison with other novel works such as KernelGAN \cite{KernelGAN}, ZSSR \cite{zeroshotsr}, DBPI \cite{DBPI}, DAN \cite{DAN}, IKC \cite{IKC}, and SRResCGAN \cite{SRResCGAN}. These SR results demonstrate that SRTGAN significantly reduces the amount of noise in the SR image and improves image clarity in comparison to other novel methods. In additon, SRTGAN can produce colours similar to the ground truth, while competing methods like IKC and KernelGAN over-boosts the colours in the generated images.

Our proposed method - SRTGAN produces SR images of better quality and with fewer noise artifacts than existing state-of-the-art methods. The quantitative assessment of several quality metrics (see Table~\ref{tab:Compare}) and the perceptual quality acquired on various datasets (see Fig.~\ref{fig:val1}-~\ref{fig:val5}) support this conclusion.\looseness=-1
\vspace{-1mm}
%Additionally, the SR results on the testing dataset of NTIRE-2020 Real SR Challenge has also been depicted in Fig. \ref{fig:test} for more intuitive comparison.
% obtained using the model without QA Network, model without Triplet Loss (Vanilla GAN Loss), and the proposed method. The original HR image is shown as Ground Truth in Fig.~\ref{fig:val1}-Fig.~\ref{fig:val2}. Looking at the SR results, one can observe that proposed method is able to outperform the other 2 methods, thereby producing better results.
% The efficacy of the proposed method has also been verified on the testing datasets of NTIRE-2020 Real-World SR Challenge for Track-1 and Track-2. In Track-1 dataset, the real-world LR images have been provided while in Track-2 the real-time data acquired using smart-phone camera are made available. We compare the SR results obtained using different methods along with the proposed method on Track-1 dataset in Fig.~\ref{fig:val2}. Similar to the earlier experiment, one can notice the superiority in performance of the proposed method as compared to same with the other existing methods. Finally, the SR results on Track-2 dataset are showb in Fig.~\ref{fig:val3} which shows that the SR results obtained using the proposed method preserves smooth regions than that of other methods.  
\section{Limitations}
\label{sec:limitations}
\vspace{-2mm}
%\kr{A discussion on known limitations will also add value to article}\ku{INCLUDED}
The proposed work obtains better results on real-world data; however, we note certain limitations as well. The network is stable only when fine-tuned for all the losses. As we can observe in Fig.~\ref{fig:val1}, the removal of the QA loss leads to undesirable outputs. Thus, fine-tuning of each loss is an expensive process. Another limitation for using the current model is that the generator and discriminator are trained in a supervised manner and hence it requires true HR-LR image pairs which can be difficult to obtain as this will need the same image to be clicked by cameras of two different resolutions. However, our work can be easily extended to unsupervised approach, as the core idea of generative modeling is to treat such unsupervised problems in a supervised manner.
% Though the proposed method produces better quality images, the core of the super-resolution problem is adding information to the image which was missing, hence, the network cannot be trusted to be exact when super resolving minute details such as architectural details, text etc. 

\section{Conclusion}
\label{sec:conclusion}
\vspace{-2mm}
We have proposed an approach to the SISR problem based on TripletGAN that fuses the novel triplet loss and no-reference quality loss along with the other conventional losses. We further modify the design of discriminator to be a patch-based discriminator for improving image quality at the scale of local image patches. The triplet loss uses both high-resolution and low-resolution images and hence, it captures the essential information required in the SR image. Applying patch-wise triplet loss improves the adversary as it allows the discriminator to better distinguish the main subject(foreground) of SR and HR images, which helps in generating images with better perceptual fidelity. Through experiments, we have demonstrated that SRTGAN can super-resolve images by a factor of $\times 4$ with improved perceptual quality than other competing methods. 
%In order to obtain an SR image of real-world LR data using deep learning approaches, generating true LR-HR pair dataset is a tedious, costly, and effort-consuming process. We propose an alternative solution using unsupervised training of deep network to generate SR images without the use of true pair LR-HR data. The proposed approach by incorporating Variational AutoEncoder (VAE) along with the discriminator network of GAN, can super-resolve the images upto a upscaling factor of $\times 4$ of original LR image size. With a newly introduced deep network based on no-reference quality assessment as a loss function, the proposed approach also enhances the perceptual fidelity of SR result. Through the experiments on NTIRE-2020 Real-world SR challenge Track-1 and Track-2 datasets, this work has demonstrated consistently superior performance for super-resolution task both in terms of objective quality metrics and perceptual quality production. In what remains, we intend to further explore alternative network architectures to mitigate the inadvertent blur and minor noise artifacts in the final SR image in future works in this direction.

\bibliographystyle{IEEEtran}
\small{\bibliography{reference}}

\end{document}